\documentclass[dvips]{acta}

\usepackage{amsmath,amssymb}

\begin{document}

\begin{center}

{\large\bf How Li and Paczy\'nski Model of Kilonova \\ Fits GW170817 Optical Counterpart}
\vskip1cm
A.~~~K~r~u~s~z~e~w~s~k~i\footnote{e-mail:  ak@astrouw.edu.pl}
\vskip3mm
Warsaw University Observatory, Al.~Ujazdowskie~4, 00-478~Warszawa, Poland
\end{center}

\Abstract{

The original Li and Paczy\'nski model of kilonova was compared with the observed bolometric optical light curve of the GW170817 electromagnetic counterpart. Perfect agreement is obtained for early observations up to 1.5 d since the time of merger.}

\section{Introduction}

Multi-messenger observations of the gravitationl waves source GW170817\break (Abbott \etal 2017) have definitely confirmed the existence of kilonova as a new kind of cosmic objects. The existence of such objects has been postulated twenty years ago by Li and Paczy\'nski (1998, hereinafter LP98).

It has been known at that time that due to the merger tidal tailes, a part of neutron stars mass is expected to be converted by r-process to radioactive elements and isotopes (Lattimer and Schramm 1974, 1976)

Considering the merger of two neutron stars or a single neutron star with a black hole LP98 omitted details of ejecting some neutron star matter out of a merging binary system. They have assumed that the sub-relativistic expanding cloud is spherically symmetric, its mass $M$ is constant with time. Its density is uniform in space while decreasing with time. They consider the interrelation between adiabatic expansion, heat generation by r-process radioactive elements and isotopes, and radiative heat losses from the cloud surface.

For an early, including the point of maximal luminosity, interval of time, they obtained an analytic solution for the bolometric luminosity dependence on time expressed in terms of Dawson function.

This paper is devoted to the comparison of LP98 model with the observational data on the GW170817 electromagnetic counterpart compiled by Waxman \etal (2017, hereinafter W17) and by Coughlin \etal (2018, hereinafter C18).

\vspace{3mm}

\section{Basic Properties of Li and Paczy\'nski Kilonova Model}

In this section  we shall present specifications of LP98 model and show the relations between the model parameters and the observational properties of the related optical transient.

An important assumption, suggested by David Spergel and adopted by LP98, is the approximate shape of the time dependence of the heat generation per gram per second $\epsilon$
\begin{equation}
\epsilon~=~\frac{fc^{2}}{t},~~~~for~~~~t_{min}\leq t \leq t_{max},~~~t_{min} \ll t_{max},
\end{equation} 
where c is the speed of light, the time $t$ is expressed in days and the energy input is parametrized by dimensionless $f$, which includes also the thermalization efficiency. The limit $t_{min}$ is supposed to be larger that a moment when the r-process has stopped to operate. The limit $t_{max}$ is set by a moment of significant decrease in the thermalization efficiency.
This expresion can be written in more general form $\epsilon \propto t^{-\alpha}$. In LP98 model it is assumed that $\alpha = 1.0$.

The average opacity is $\kappa$. The opacity caused by electron scattering is $\kappa_{e} = 0.2~cm^{2}g^{-1}$ and it is used as the reference value of opacity.

The time $t_{c}$ at which the expanding sphere became optically thin is
\begin{equation}
t_{c}~=~\left (\frac{3\kappa M}{4\pi V^{2}} \right )^{1/2},
\end{equation}
where $V=~\beta\times c$ is the expansion speed at the surface and $M$ is mass of the ejected matter.

Using Eq.(1), LP98 have obtained following approximate expression for the time dependence of the bolometric luminosity $L$
\begin{equation}
L~\approx ~L_{0}~ \sqrt{\frac{8\beta}{3}}~~Y\left(\sqrt{\frac{3}{8\beta}}~\tau\right),
\end{equation}
where $L_{0}= fMc^{2}/(4\beta t_{c})$, ~$\tau=t/t_{c}$, ~Dawson function $Y$ is
\begin{equation}
Y(x)~=~e^{-x^{2}} \int_{0}^{x} e^{s^{2}}ds.
\end{equation}

When building a simple model for a specific object one needs to know 5 parameters $\alpha, ~\beta, ~\kappa, ~f$ and $M$. After fixing the value of $\alpha$, approximate values of the time from the beginning of expansion to the peak luminosity $t_{m}$, the peak luminosity $L_{m}$, and the effective temperature at the peak luminosity $T_{eff,\hspace{0.5mm}m}$ can be expressed as functions of the remaining four parameters
\begin{equation}
t_{m}~\approx ~0.98~d \left (\frac{M}{0.01 M_{\odot}} \right )^{1/2} \left (\frac{3V}{c} \right )^{-1/2} \left (\frac{\kappa}{\kappa_{e}} \right )^{1/2} ,
\end{equation}
\begin{equation}
L_{m} ~\approx ~2.1 \times 10^{44}~ergs/s \left (\frac{f}{0.001} \right ) \left ( \frac {M}{0.01 M_{\odot}} \right )^{1/2} \left (\frac {3V}{c} \right )^{1/2} \left ( \frac{\kappa}{\kappa_{e}} \right )^{-1/2} ,
\end{equation}
\vspace{0.8mm}
\begin{equation}
T_{eff,\hspace{0.5mm}m}~ \approx ~2.5 \times 10^{4}~K \left (\frac{f}{0.001} \right)^{1/4} \left (\frac{M}{0.01~M_{\odot}} \right )^{-1/8} \left ( \frac{3V}{c} \right )^{-1/8} \left (\frac{\kappa}{\kappa_{e}} \right )^{-3/8} .
\end{equation}
\vspace{0.6mm}

LP98 do not propose any strict predictions for the parameters $\beta, ~\kappa, ~f$ and $M$. Only the plausible reference values of these parameters are used in Eqs.(5)-(7). 

LP98 noticed that in the optically thin region the produced and emitted energies are equal and consequently
\begin{equation}
L(t)~ \approx ~M \times \epsilon(t)~=~\frac{Mfc^{2}}{t}.
\end{equation}

In addition, LP98 give an equation that serves as a prediction. They give an estimate of time needed for luminosity to drop by factor of 3 from its peak value
\begin{equation}
\Delta t~ = ~2.27 \times t_{m}~ d. 
\end{equation}

\section{Li and Paczy\'nski Model Fitted to the GW170817 Optical Counterpart}

The photometric observations of the GW170817 optical counterpart were published in several papers. The compilation of W17 seems to suit best for our comparison. We have used the integrated luminosities $L_{int}$ from Table 3 of their's paper. These measurements of bolometric luminosity have been confronted with Eq.(3). Using the Levenberg-Marquart algorithm (Press \etal 2007) applied to the 6 earliest data points we obtain the following values of $t_{m}$ and $L_{m}$
\begin{equation} 
t_{m}~ = ~(0.600 \pm 0.008)~d,~~~~~~~~~~~L_{m}~ = ~(7.92 \pm 0.11)\times 10^{41}~ergs/s.
\end{equation}
We can read from the same Table 3 in W17 the value of temperature at the maximum luminosity
\begin{equation}
T_{eff,\hspace{0.5mm}m} = ~(10809 \pm 712)~K,
\end{equation}
so we have estimates for the left hand sides of Eqs.(5)-(7).

Fig. 1 presents with black symbols the comparison of the measurements of bolometric luminosity with LP98 model fitted to the six earliest data points. The blue square correspons to the test point suggested by LP98. The earliest six observations fit surprisingly well to the used by LP98 Dawson function. Next four observations seem to lie along a straight line whose slope agrees with Eq.(8).

We shall try to understand the general picture as follows: The basic quantity is a total cloud heat. In LP98 model the total cloud heat is fed by the radioactive heat input and it is a source of the cloud bolometric luminosity. LP98 model describes the interrelations between these three components in the time range between 0.5 and 1.5 days. There should exist in that time range a moment $t_{equ}$, when originally increasing total cloud heat attains its maximum value. At that moment the emited bolometric luminosity is equal to the energy produced by the radioactive process. The time dependences of both these quantities follow Eq.(8). The location of this moment can be obtained by requiring the local fulfilment of the Eq.(8) soon after the bolometric luminosity maximum. The result is
\begin{equation}
t_{equ}~ = ~0.97563~~d,~~~~~~~~~~~~~~~~~L_{equ}~= ~6.25943\times 10^{41}~ergs/s,
\end{equation}
and it is marked in Fig.1 with a star symbol. 

\begin{figure}[htb]
\includegraphics*[width=12.5cm,bb= 6 0 774 588]{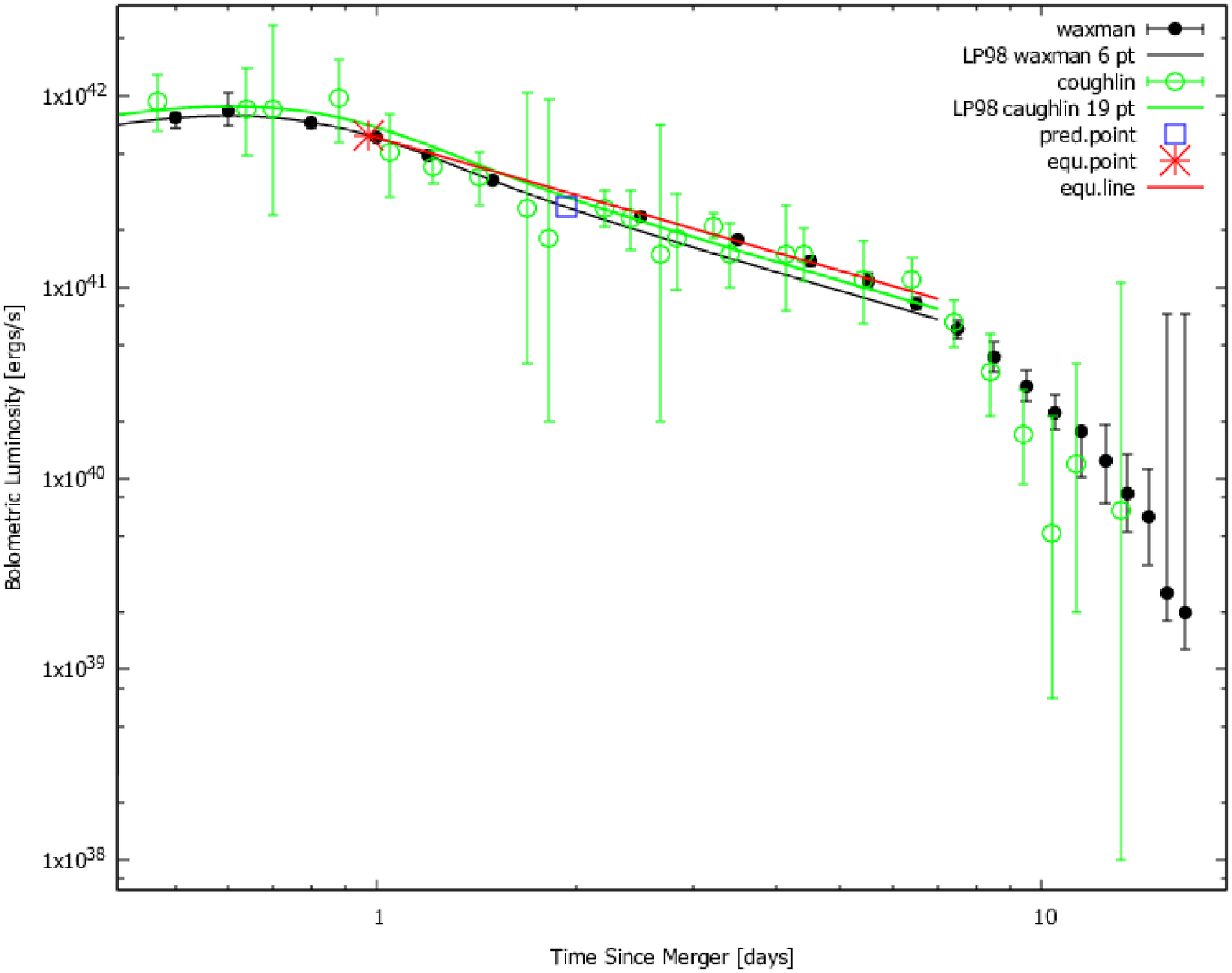}
\FigCap{The black points with corresponding error bars are from Table 3 in W17. The black curve presents LP98 model fitted to the six earliest black points. The green circles with corresponding error bars are from Table 2 in C18. The green curve presents LP98 model fitted to the 19 earliest green circles. The blue square shows the predicted by LP98 time when the luminosity is decreased threefold from the luminosity maximum. The red star points to the moment when the total cloud heat attains its maximum value. The red line has a starting point at the red star and its slope is fixed as $-1.0$.}
\end{figure}

Let us assume, that a time interval between 2.5 and 5.5 days since merger is contained in the optically thin region. Therefore in that interval, all the produced heat is equal to the heat radiated. 
Starting from point $t_{equ}$, we shall draw in Fig. 1 a line with its slope equal to $-1.0$ and this line shows very good agreement with four data points inside time interval $[2.5,5.5]$, what seems to confirm our interpretation of the point $t_{equ}$ and also our assumption that 2.5-5.5 days is inside the optically thin time region. 

We shall look closely on the procedure used by W17 at obtaing the luminosity data points $L_{int}$ listed in theirs Table 3. For each of photometric filters the available measurements have been subjected to polynomial smoothing over time. The results of this smoothing were used to calculate the bolometric luminosity as a function of time. This was done by means of trapezoidal integration $L_{int}$ and also by fitting a blackbody $L_{bb}$. The trapezoidal integration data points were adapted by W17 as the best estimate of the photometric data. Therefore they are used in this paper. We have used the errors read out from Table 3 in W17 for weighting individual data points only. The formal errors quoted in Eq.(10) are certainly underestimated. For getting more reliable error estimates we should repeat fitting LP98 model to another data set with statistically independent observational errors. A compilation by C18 seems to fulfill this requirement and it is plotted in Fig. 1 with green circles. We can see in Fig. 1 that within the time interval $0.4 - 7.0$ days there are no easily visible systematic differences between these two compilations.

We fit LP98 model to 19 C18 earliest data points up to 7 days after the merger where, according to W17, the spectra start to deviate significantly from blackbody. The resulting bolometric luminosity curve is pictured in Fig. 1 with green line.
\begin{equation} 
t_{m}~ = ~(0.604 \pm 0.013)~d,~~~~~~~~~~~L_{m}~ = ~(8.90 \pm 0.33)\times 10^{41}~ergs/s,
\end{equation}
with correlation coefficient between $t_{m}$ and $L_{m}$ equal to $-0.34$.  

Comparing estimates of $t_{m}$ and $L_{m}$ from Eqs.(10) and (13) based respectively on M17 and C18 compilations we can see that values of $t_{m}$ are almost identical. Values of $L_{m}$ differ by $3\sigma$. This difference can be at least partly explained by using only the data corresponding to the optically thick region when using W17 data while in the case of C18 data from both thick and thin regions were used.

The basic results of LP98 model are contained in Eqs.(5)-(7) which describe the model with help of 4 parameters $V, ~M, ~\kappa$ and $f$ while we have only three equations. However we can solve these equations with a reduced set of unknows, namely $V$,~$M \times f$ and $M\times \kappa$ or $f / \kappa$. 
\begin{equation}
\left (\frac{3V}{c} \right )~ \approx~ \left ( \frac{L_{m}}{2.1\times10^{44}~ergs~s^{-1}} \right)^{1/2} \left (\frac{T_{eff,~m}}{2.5\times 10^{4}~K} \right )^{-2} \left (\frac{t_{m}}{0.98~d} \right )^{-1},
\end{equation}
\begin{equation}
\left (\frac{M}{0.01M_{\odot}} \right ) \left (\frac{f}{0.001} \right )~ \approx ~\left ( \frac{L_{m}}{2.1\times10^{44}~ergs~s^{-1}} \right)  \left (\frac{t_{m}}{0.98~d} \right ),
\end{equation}
and
\begin{equation}
\left (\frac{M}{0.01M_{\odot}} \right ) \left (\frac{\kappa}{\kappa_{e}} \right )~ \approx ~\left (\frac{L_{m}}{2.1\times10^{44}~ergs~s^{-1}} \right )^{1/2} \left (\frac{T_{eff,~m}}{2.5\times 10^{4}~K} \right )^{-2} \left (\frac{t_{m}}{0.98~d} \right )~,
\end{equation}
or
\begin{equation}
\left (\frac{f}{0.001} \right ) \left (\frac{\kappa}{\kappa_{e}} \right )^{-1} \approx ~\left ( \frac{L_{m}}{2.1\times10^{44}~ergs~s^{-1}} \right)^{1/2}  \left (\frac{T_{eff,~m}}{2.5\times10^4~K} \right )^{2}.
\end{equation}
\vspace{0.6mm}

Looking closer on Eq.(17) we can notice that the left hand side of this equation is based on the atomic data, while we have the observational data on the right hand side. Therefore we can judge with it on the quality of used atomic data.

Using Eqs.(11) and (13) we obtain
\begin{equation}
V~\approx ~(0.188\pm 0.026)~c,
\end{equation}
\begin{equation}
M\times f~ \approx~ (2.61 \pm 0.10)\times10^{-8}~M_{\odot},
\end{equation}
and
\begin{equation}
f / \kappa~ \approx~(6.1\pm 0.8)\times 10^{-5}~cm^{-2}~g.
\end{equation}
or
\begin{equation}
M \times \kappa~ \approx~(4.3 \pm 0.6 )\times 10^{-4}~M_{\odot}~cm^{2}~g^{-1}.
\end{equation}
\vspace{-0.6mm}

The lower limit for $\kappa$ may be set as
\begin{equation}
\kappa ~\gtrapprox ~0.1~ cm^{2} g^{-1},
\end{equation}
and that makes it possible to get the following inequalities
\begin{equation}
M~ \lessapprox ~(4.3 \pm 0.6) \times 10^{-3}~M{\odot} ,
\end{equation}
\begin{equation}
f~ \gtrapprox ~(6.1 \pm 0.8) \times 10^{-6}.
\end{equation}

It has been pointed out by M10 that the used in Eqs.(6) and (7) reference value of $f$ equal to $0.001$ is unrealistically large. However, the plausible values of $f$, as they are pictured in Fig. 2 of LP98, range from $10^{-5}$ to $10^{-3}$ and the lower limit $10^{-5}$ is only little larger that the lower limit for $f$ according to Eq.(24).  

It would be interesting to compare results obtained with the same set of data in C18 and in this paper. Comparing our Fig. 1 with Fig. 1 in C18 we can see that theirs time of maximum luminosity $t_{m}$ is smaller that 10 hr, while we have obtained $t_{m} = (14.5\pm 0.3)$ hr. Another striking difference is seen in estimates of the ejecta mass $M$. We can conclude from Eq.(23) that $M\ll 0.01M_{\odot}$, while in C18 the single component fit results with $M \approx ~0.05M_{\odot}$ and two component fit would require $M\approx ~0.055M_{\odot}$.

Any of the used approaches, analytic or Monte Carlo or both of them, can suffer from unaccounted for systematic errors or limitations. Optical observations of new such objects covering also first hours after the merger should help to settle this question.

\section{Conclusions}

LP98 is a spherically symmetric, homogeneous, single component model.
In spite of being simple, LP98 model has presented a new look on the observable remnants of the neutron star merging with black hole or another neutron star. They have realized that the presence of r-process nuclei will be followed by a continuous nuclear explosion in expanding with sub-relativistic velocity cloud, which after being enlarged by factor $\approx 10^{8}$ can be sufficiently big and bright to be observed in the optical spectral region. All the following papers describing this kind of explosion have their roots in Li and Paczy\'nski model. The name for this model ``kilonova'' has been coined by Vah\'e Petrosian (M10).

LP98 have noticed, assuming correctness of Eq.(1), that the overall heat balance leads to the generic equation for Dawson function. This has resulted in obtaining a simple analytical solution for the earliest and also brightest part of the luminosity-time dependence. They have noticed also that later on, in the optically thin region, the luminosity may follow Eq.(8), forming a straight line on a log-log luminosity vs time diagram. Fig. 1 shows surprisingly good, though possibly fortuitous, agreement of these features with observations.

Such a fine agreement between model and observations need confirmation from data on similar objects. After such confirmation LP98 model may be useful at early selection of kilonovae among optical transits. That may help to find orphan kilonovae at the absence of gravitational and/or gamma ray counterparts. It may also suggest early approximate limits for the object parameters what can be useful at constructing more sophisticated models.

\Acknow{I thank the anonymous referee for thoughtful report that helped to improve this paper. I am indebted to Li-Xin Li, Micha{\l} R\'o\.zyczka and Micha{\l} Szy\-ma\'nski for reading and commenting on the manuscript.}

\end{document}